\documentclass[pra,aps,twocolumn,eqsecnum,groupedaddress,nofootinbib,superscriptaddress]{revtex4}
\usepackage{graphicx}
\usepackage{bm}
\usepackage{epsfig}
\usepackage{amssymb}
\usepackage{amsfonts}
\usepackage{amsmath}
\usepackage{color}
\usepackage{xcolor}
\usepackage{epstopdf}
\usepackage{hyperref}
\usepackage{float}
\usepackage{appendix}
\usepackage{wasysym}
\restylefloat{table}
\usepackage{bibentry}
\usepackage{multirow}
\usepackage[caption=false]{subfig}
\usepackage{braket}
\usepackage{bbm}
\newcommand{\ba}{\begin{eqnarray}}
\usepackage{manfnt}
\newcommand{\ea}{\end{eqnarray}}
\newcommand{\bd}{\begin{displaymath}}

\newcommand{\nn}{\nonumber \\}

\DeclareGraphicsExtensions{.pdf,.png,.jpg}

\begin{document}

\title{Hybrid rank-1 and rank-2 U(1) lattice gauge theory, the F3 model, \\ and its effective field theory}

\author{Jintae Kim}
\affiliation{Department of Physics, Sungkyunkwan University, Suwon 16419, Korea}
\author{Yun-Tak Oh}
\affiliation{Department of Physics, Korea Advanced Institute of Science and Technology, Daejeon 34141, Republic of Korea}
\affiliation{Division of Display and Semiconductor Physics, Korea University, Sejong 30019, Korea}
\author{Jung Hoon Han}
\email[Electronic address:$~~$]{hanjemme@gmail.com}
\affiliation{Department of Physics, Sungkyunkwan University, Suwon 16419, Korea}

\date{\today}
\begin{abstract}
A number of exactly solvable spin models, including the Kitaev toric code in two and three dimensions and the X-cube model in three dimensions, can be related to their respective parent lattice gauge theories (LGT) through the mathematical process of `Higgsing'. Field theories of the low-energy excitations of these spin models can be developed subsequently, building upon the symmetry of the parent LGTs. Recently, two of the present authors proposed a variant of the three-dimensional toric code which we now call the F3 model, whose elementary excitations consist of freeon and fluxon excitations of the three-dimensional toric code and fracton excitations of the X-cube model. In this work, we identify the parent LGT of the F3 model as the hybrid rank-1 and rank-2 U(1) LGT, and develop the corresponding field theory. The resulting Lagrangian of the F3 model is that of a three-dimensional toric code with an extra term, which ties the dynamics of fractons to that of fluxons. The matter part of the effective action for the F3 model can be derived as well, by carefully keeping track of the gauge invariance of the F3 model. Hydrodynamic equations of motion of the quasiparticles are derived, which properly reflect their constrained dynamics. Finally, we present a tight-binding model for the quasiparticle motion in the F3 model.
\end{abstract}
\maketitle
\section{Introduction}

Exactly solvable spin Hamiltonians such as toric codes (TC) in two and three dimensions~\cite{kitaev03, simon13, freedman01} and the X-cube model in three dimensions~\cite{vijay16, slagle17, hermele19} have been studied extensively. 
The multi-spin interactions that define various exactly solvable spin models seem at first novel, but in fact have well-defined roots in their corresponding parent lattice gauge theories (LGT)~\cite{hermele18b, barkeshli18,lewenstein13}. The mutually commuting operators in the LGT - also known as generators - become, through Higgsing, the stabilizer operators in the spin Hamiltonian. Various rank-1 and rank-2 U(1) LGTs and their Higgsed spin models have been analyzed, and it was shown that the parent theory of the TC is the rank-1 U(1) LGT, while that of the X-cube model is the rank-2 U(1) LGT~\cite{barkeshli18}. The elementary excitations of the toric code are the electric and magnetic quasiparticles exhibiting mutual anyonic statistics, while those of the X-cube model are the fractons and lineons. Fractons are immobile {\it but} deconfined in the sense that a single fracton cannot move without extra energy cost, but a four-fracton cluster can be expanded arbitrarily without extra energy. Lineons are subdimensional quasiparticles that can only move freely along a specific axis and are also deconfined~\cite{vijay16}. 

In an attempt to extend the realm of exactly solvable spin models, two of the present authors proposed a model~\cite{kim21} exhibiting features akin to those of the three-dimensional toric code (3DTC) as well as the X-cube model. The new model is obtained from the 3DTC by including an additional twelve-spin interaction which, in the context of the X-cube model, acts as a fracton operator. Excitations in this new model consist, at first sight, of the two kinds of elementary excitations of the 3DTC, augmented by the fracton excitation similar to that in the X-cube model. In a previous publication~\cite{kim21} these excitations were dubbed freeons, fluxons, and fractons, respectively, and hence we will call the overall model `F3 model'. It was also shown~\cite{kim21} that the model is exactly solvable only for the $\mathbb{Z}_3$ Hilbert space at the links of the lattice, and the ground state degeneracy (GSD) of the model is $3^3$, same as the GSD of the $\mathbb{Z}_3$ 3DTC. In other words, the F3 model remains in the same phase as the 3DTC despite the addition of the local fracton operators. A physical way to understand this is that fractons that are deconfined in the X-cube model become confined in the F3 model because fluxons act as `gluons' between fractons with energy that increases linearly with separation. The confinement of fractons in the F3 model is an indication that the low-energy field theory of the F3 model is more likely to resemble that of 3DTC than that of the X-cube model. We will show that this is indeed the case.

Neither the parent LGT of the F3 model nor its low-energy field theory was properly understood in the previous work~\cite{kim21}. We show in this paper that the F3 model is the descendant of the hybrid rank-1 and rank-2 U(1) LGTs. 
The field theory of the F3 model is constructed accordingly. In Sec. \ref{sec:2} we review the rank-1 and rank-2 U(1) LGTs, their descendant exactly solvable spin models, and the corresponding effective field theories (EFTs). Although the discussions in this section overlap with several works in the past, we find it useful to collect known results and present them in a unified fashion, thus paving the way for the development of EFT of the F3 model. Some new insights as well as new results about the toric code and X-cube model in three dimensions are obtained. In Sec. \ref{sec:3} we adopt the strategies developed in Sec. \ref{sec:2} and apply them to the F3 model. The parent LGT of the F3 model is properly identified and is shown to lead to the F3 stabilizer Hamiltonian upon Higgsing. The gauge transformation properties of the various gauge fields are clarified and subsequently exploited to derive the field theories of the matter fields and the gauge fields pertinent to the F3 model. Section \ref{sec:5} gives the summary. Several technical themes are delegated to the appendices. 

\section{Review of Rank-1 and Rank-2 U(1) LGT and their descendants}
\label{sec:2}

We review the rank-1 and rank-2 U(1) LGT, their descendant exactly solvable spin models, and the corresponding EFTs. 

\subsection{Rank-1 U(1) LGT and its descendants}
\label{subsec:2A}

The rank-1 U(1) LGT has the vector gauge fields $A_i^a$ residing on the $(i, i+\hat{a})$ links of the lattice, where  $a=x,y,z$. The canonically conjugate electric fields $E_i^a$ satisfy the commutation relation $[A_i^a,E_j^b]=i\delta_{ij}\delta_{ab}$. The gauge transformation of $A_i^a$ and $E_i^a$ is generated as
\begin{align}
A_i^a \rightarrow U^\dag_A A_i^a U_A &= A_i^a + f_{i+\hat{a}}-f_{i}  , \nn 
& \sim A_i^a+\partial_a f_i\nn
E_i^a \rightarrow U_E^\dag E_i^a U_E &= E_i^a +\sum_{bc} \epsilon_{abc} (g_i^c-g_{i-\hat{b}}^c)\nn
&\sim E_i^a +\sum_{bc} \epsilon_{abc} \partial_b g_i^c,
\end{align}
where the unitary matrices $U_A$ and $U_E$ are
\begin{align}
U_A = e^{ i \sum_{j} f_j ({\bm \nabla} \cdot {\bm E})_j } , ~~~
U_E = e^{i \sum_{j} {\bm g}_j \cdot {\bm B}_j } 
\label{eq:gt}
\end{align}
for arbitrary functions $f_i$ and $g_i^a$. The spatial indices $a$, $b$, $c$ run over $x,y,z$. The lattice divergence of the electric field and the magnetic field operator, called generators, are defined by 
\begin{align}
({\bm \nabla} \cdot {\bm E})_i &\equiv \sum_{a}  (E_i^a-E_{i-\hat{a}}^a) \sim \sum_{a}  \partial_a E_i^a\nn
B_i^a&\equiv \sum_{bc} \epsilon_{abc}( A_i^b-A_{i+\hat{c}}^b) \sim \sum_{bc} \epsilon_{abc} \partial_b A_i^c .
\label{eq:div}
\end{align}
Importantly, we have the commutativity of the two generators
\begin{align}
\left[ \left( {\bm \nabla} \cdot {\bm E}\right)_i, B_j^a \right]=0 ,\label{eq:commute}
\end{align}
following from their definitions.

The Hamiltonian produced from Higgsing such rank-1 U(1) LGT is precisely the 3DTC. The Higgsing is given by
\begin{align}
X_{a,i} &= \exp (- 2\pi i E_i^{a}/p )  , &Z_{a,i} &= \exp (i A_i^{a} )   \label{eq:discrete}
\end{align}
for $a=1,~2,~3$ corresponding to $a=x,~y,~z$, respectively. The generalized Pauli operators $X, Z$ obey the commutation algebra 
\begin{align} ZX = \omega XZ  ~~ ( \omega= e^{2\pi i/p} ) \end{align} 
for integer $p \ge 2$. 
%
%
The $\mathbb{Z}_p$ 3DTC~\cite{simon13} is defined as
\begin{align} 
H_{\rm 3DTC} & = -\sum_i (\mathcal{A}_i + \mathcal{C}_i^x + \mathcal{C}_i^y +  \mathcal{C}_i^z ),  \label{eq:2D3DTC}
\end{align} 
where
\begin{align}
\mathcal{A}_i &= \frac{1}{p} \sum_{j =0}^{p-1} (a_i )^j ~~~~~~~\mathcal{C}_i^a = \frac{1}{p} \sum_{j =0}^{p-1} (c_i^a )^j\nn
a_i &= X_{1,i} X_{1,i-\hat{x}}^{-1} X_{2,i} X_{2,i-\hat{y}}^{-1} X_{3,i} X_{3,i-\hat{z}}^{-1}\nn
c_i^x  &= Z_{2,i} Z_{3,i+\hat{y}} Z^{-1}_{2,i+\hat{z}} Z^{-1}_{3,i}\nn
c_i^y  &= Z_{3,i} Z_{1,i+\hat{z}} Z^{-1}_{3,i+\hat{x}} Z^{-1}_{1,i}\nn
c_i^z  &= Z_{1,i} Z_{2,i+\hat{x}} Z^{-1}_{1,i+\hat{y}} Z^{-1}_{2,i}.
\label{eq:toricab}
\end{align}
Electric quasiparticle is the excitation of $A_i$ while magnetic quasiparticles labeled as $m_a$ ($a=x,y,z$) are the excitations of $C_i^a$.

The EFT for 3DTC is the well-known BF theory, but the treatment of the EFT for the matter fields, i.e., electric and magnetic excitations, is not given with equal care in the literature. Here we derive the EFTs of both matter and gauge fields for the 3DTC and derive the corresponding continuity equations. Furthermore, a tight-binding realization of the $e$-particle and $m$-particle is given by invoking the gauge principle.

The effective action of the $e$ quasiparticles is the usual one,  
\begin{align}
\mathcal{L}_{e}& =i\psi_e^\dag \partial_t \psi_e- \frac{1}{2m_e} | {\bm D}_e \psi_e |^2 \label{eq:TCL}, \end{align}
with ${\bm D}_e \psi_e = ({\bm \nabla} - i {\bm A})\psi_e$. Its form is dictated by the requirement of invariance under the gauge transformation $\psi_e \rightarrow e^{if} \psi_e$ and ${\bm A} \rightarrow {\bm A} + {\bm \nabla} f$. The mass $m_e$ is introduced by hand. The EFT of the $m_a$ quasiparticles is more involved,  
\begin{align}
\mathcal{L}_{m}& =i{\bm \psi}_m^\dag \cdot \partial_t {\bm \psi}_m - \frac{1}{2m_m} \sum_{a < b} | {\bm D}_m^{ab} \psi_m^a \psi_m^b |^2,
\end{align}
with ${\bm \psi}_m = (\psi_m^x, \psi_m^y, \psi_m^z)$ representing the three species of magnetic quasiparticles. The covariant derivative $D_m^{ab}$ must be chosen to comply with the rule of gauge invariance, as follows (no sum on $a,b$):
\begin{align} 
{\bm D}_m^{ab} \psi_m^a \psi_m^b & =\psi_m^a\partial_a \psi_m^b-\psi_m^b\partial_b \psi_m^a\nn 
& -i\sum_c \epsilon_{abc} E^c\psi_m^a\psi_m^b. \label{eq:3DTC-cov-der} \end{align} 
The gauge invariance of the action under $\psi_m^a \rightarrow e^{ig^a} \psi_m^a$ and $E^a \rightarrow E^a + \sum_{bc} \epsilon_{abc} \partial_b g^c$ can be readily checked. The action is quartic in the fields and implies that the quasiparticles are inherently interacting. It is well-known that magnetic excitations in the 3DTC always form loops~\cite{simon13}, which one can now understand as a consequence of the strongly interacting nature of the magnetic quasiparticles.

Some conservation laws of electric and magnetic quasiparticles follow from the matter action: 
\begin{align}
&\partial_t \rho^e + {\bm \nabla} \cdot {\bm j}^e =  0 , \nn
&\partial_t \rho^{m_a} + \sum_{b} \partial_b [j^m]^{ab}  =  0 .\label{eq:3dtccl}  
\end{align}
Explicit expressions of the charge and current densities in terms of the matter fields can be found in the Appendix \ref{asec:B}. The electric part of the conservation law is nothing out of the ordinary. For the magnetic part, a vector of charges $\bm \rho^m=(\rho^{m_x}, \rho^{m_y}, \rho^{m_z})$ and the corresponding antisymmetric tensor current $[j^m]^{ab}$ give rise to an unusual form of continuity equation. In fact, one can write the current tensor in a vector form ${\bm j} = ([j^m]^{yz}, [j^m]^{zx}, [j^m]^{xy})$, introduce the vector charge density ${\bm \rho}^m = (\rho^{m_x} , \rho^{m_y} , \rho^{m_z})$, and simplify the continuity equation of the magnetic particles:
\begin{align}
    \partial_t {\bm \rho}^m + {\bm \nabla} \times {\bm j} = 0 . \label{eq:3DTC-curr-cons-simplified} 
\end{align}
A natural corollary of this equation is that $\partial_t ({\bm \nabla} \cdot {\bm \rho}^m ) = 0$, consistent with the fact that magnetic excitations in 3DTC only form closed loops. The construction of EFTs for the matter parts of the 3DTC and the corresponding continuity equations are not treated in the literature, and we presented them in some detail here. 

The Lagrangian reflecting the commutation relation of the gauge fields $( A_i^a , E_j^b )$ and the conservation laws in Eq. (\ref{eq:3dtccl}) is the three-dimensional BF theory. The action can be cast in the four-vector continuum notation ($\epsilon_{0123} = +1$)~\cite{hansson04}:
\begin{align}
\mathcal{L}_{\text{BF}} = \frac{p}{2\pi} \epsilon_{\alpha\beta\mu\nu} \frac{1}{2} E^{\alpha\beta} \partial_\mu A^\nu - A \cdot j^e - \frac{1}{2} E^{\alpha\beta} {[j^m]}^{\alpha\beta}, \label{eq:EFT-for-3DTC}
\end{align}
where $A$ and $j^{e}$ are four-vectors, $A = (A^0, {\bm A} )$, $j^e = (\rho^e , {\bm j}^e )$, and $E^{\alpha\beta}$ and ${[j^m]}^{\alpha\beta}$ are antisymmetric tensors whose definitions are written out in Appendix \ref{asec:B}. The four-current densities of electric and magnetic quasiparticles are denoted $j^e$ and $[j^m]^{\mu\nu}$, respectively. The commutation relation and the gauge transformations of the $A^a$ and $E^{ab}$ are
\begin{align}
[A^a({\bm r}),&E^{bc}({\bm r'})]=\frac{2\pi i}{p} \epsilon_{abc} \delta^3({\bm r}-{\bm r'})\nn
{\bm A} ({\bm r})&\rightarrow U_A^\dag  {\bm A} ({\bm r}) U_A ={\bm A}({\bm r})+{\bm \nabla} f ({\bm r})\nn
E^{ab} ({\bm r})&\rightarrow U_E^\dag E^{ab} ({\bm r}) U_E =E^{ab} ({\bm r})-\partial_a g^b ({\bm r})+\partial_b g^a ({\bm r}),
\end{align}
and the gauge transformations of the temporal components $A^0$ and $E^{a0}$ are
\begin{align}
A^0(\bm r) &\rightarrow A^0(\bm r)+\partial_t f(\bm r)\nn
E^{a0}(\bm r) &\rightarrow E^{a0}(\bm r)-\partial_a g^0(\bm r)+\partial_t g^a(\bm r).
\end{align}
Accordingly, $A \cdot j^e$ and $\frac{1}{2} E^{\alpha\beta} {[j^m]}^{\alpha\beta}$ transform under the gauge transformations as 
\begin{align}
A \cdot j^e &\rightarrow A \cdot j^e+\sum_\mu \partial_\mu f [j^e]^\mu\nn
\frac{1}{2} E^{\alpha\beta} {[j^m]}^{\alpha\beta} &\rightarrow \frac{1}{2} E^{\alpha\beta} {[j^m]}^{\alpha\beta} + \frac{1}{2}(-\partial_\alpha g^\beta +\partial_\beta g^\alpha)[j^m]^{\alpha \beta}.\label{eq:cld3tc}
\end{align}
Applying integration by parts to the two equations in Eq. (\ref{eq:cld3tc}) leads to the conservation laws of $e$ and $m$ quasiparticles as well as the identity $\bm \nabla \cdot \bm \rho^m=0$. The BF action for the 3DTC is well-known, but the approach that starts from the explicit construction of the matter Lagrangian followed by the identification of various conservation laws is new, and we outline the derivation in Appendix \ref{asec:B}.

We can construct a lattice version of the matter field action of the 3DTC by following the gauge principle. For the electric quasiparticle, the usual Peierls substitution
\begin{align}
[\psi_e]_{i+\hat{a}}^\dag [\psi_e]_i e^{iA^a_i} \label{eq:lelec}
\end{align}
yields gauge-invariant hopping of the electric operator $\psi_e$. On the other hand, the gauge transformations of the ${\bm E}$ field and magnetic quasiparticle operators ${\bm \psi}_m = (\psi_m^x , \psi_m^y , \psi_m^z )$ on the lattice are given by
\begin{align}
E_i^{ab} & \rightarrow E_i^{ab} -g^b_{i+\hat{a}}+g^b_{i}+g^a_{i+\hat{b}}-g^a_{i}, \nn 
\hspace{0cm}[{\bm \psi}_m]_i  & \rightarrow (e^{ig^x_i} [\psi_m^x]_i , e^{ig^y_i} [\psi_m^y]_i , e^{ig^z_i} [\psi_m^z]_i ) ,
\end{align}
and the corresponding gauge-invariant hopping terms must take the form 
\begin{align}
& [\psi_m^x]_{i+\hat{y}}^\dag [\psi_m^y]_{i}^\dag [\psi_m^x]_{i} [\psi_m^y]_{i+\hat{x}} e^{iE^{xy}_i} , \nn 
& [\psi_m^y]_{i+\hat{z}}^\dag [\psi_m^z]_{i}^\dag [\psi_m^y]_{i} [\psi_m^z]_{i+\hat{y}} e^{iE^{yz}_i} , \nn
& [\psi_m^z]_{i+\hat{x}}^\dag [\psi_m^x]_{i}^\dag [\psi_m^z]_{i} [\psi_m^x]_{i+\hat{z}} e^{iE^{zx}_i}  \label{eq:2.10} 
\end{align}
and their Hermitian conjugates. No single-particle hopping is allowed for magnetic particles; only their correlated hopping involving two species of $m$ quasiparticles at a time is allowed under the gauge principle. 

\subsection{Rank-2 U(1) LGT and its descendants}


The particular kind of rank-2 U(1) LGT we are interested in utilizes symmetric rank-2 gauge fields $A_{i}^{ab} = A_{i}^{ba}$ $(a,b=x,y,z)$ obeying the hollowness condition $A_i^{aa} = 0$ (no sum on $a$) for the diagonal components. The conjugate fields $E_i^{ab}$ satisfying the commutation $[A_i^{ab},E_j^{ab}]=i\delta_{ij}$ share the same symmetry  and hollowness properties: $E_i^{ba} = E_i^{ab}$ and $E_i^{aa} = 0$~\cite{barkeshli18}.
The transformation rule for the gauge fields are
\begin{align}
A_i^{ab} & \rightarrow U^\dag_A A_i^{ab} U_A  \nn 
& = A^{ab}_i - f_{i-\hat{a}-\hat{b}}+f_{i-\hat{a}}+f_{i-\hat{b}}-f_{i} \nn 
& \sim A^{ab}_i - \partial_a \partial_b f_i , \nn 
{E}_i^{ab} & \rightarrow U_E^\dag {E}_i^{ab} U_E  \nn 
& = {E}_i^{ab}  + \sum_{c} \epsilon_{abc} (g^b_{i+\hat{c}}-g_i^b-g^a_{i+\hat{c}}+g_i^a)\nn
& \sim {E}_i^{ab}  + \sum_{c} \epsilon_{abc} (\partial_c g^b_{i}-\partial_c g^a_{i}),
\label{eq:gt2}
\end{align}
where the unitary operators are defined by
\begin{align}
U_A = e^{ i \sum_j f_j ({\bf DE} )_j } , ~~ U_E = e^{ i \sum_{j} {\bm g}_j \cdot {\bm B}_j } 
\end{align}
for arbitrary functions $f_i$ and $\bm {g}_i$. Generators of the gauge transformations are  
\begin{align}
({\bf D E})_i & \equiv E_{i+\hat{x}+\hat{y}}^{xy}-E_{i+\hat{x}}^{xy}-E_{i+\hat{y}}^{xy}+E_{i}^{xy}\nn
&+E_{i+\hat{y}+\hat{z}}^{yz}-E_{i+\hat{y}}^{yz}-E_{i+\hat{z}}^{yz}+E_{i}^{yz}\nn
&+E_{i+\hat{x}+\hat{z}}^{xz}-E_{i+\hat{x}}^{xz}-E_{i+\hat{z}}^{xz}+E_{i}^{xz} \nn 
& \sim \partial_x \partial_y E_i^{xy} + \partial_y \partial_z E^{yz}_i + \partial_z \partial_x E_i^{xz} , \nn 
B_i^a & \equiv \sum_{bc} \epsilon_{abc} (A_{i}^{ba}-A_{i-\hat{c}}^{ba})\nn
&\sim \sum_{bc} \epsilon_{abc} \partial_c A_{i}^{ba}. 
\label{eq:R2-Gauss-law} 
\end{align}
One can easily prove the commutativity $[({\bf DE})_i ,  B_j^a ] = 0$. 

The Hamiltonian produced from Higgsing the rank-2 U(1) LGT is the X-cube model. Before introducing the Higgsing formula, we need to redefine the notation of tensor fields. There are, in fact, only three independent components in the tensors $A^{ab}$ and $E^{ab}$ and they can be mapped to vector fields 
\begin{align} 
(A_i^{yz}, A_i^{xz}, A_i^{xy}) & \rightarrow (A_{i}^x , A_{i}^y , A_{i}^z ) \nn 
(E_i^{yz}, E_i^{xz}, E_i^{xy}) & \rightarrow (E_{i}^x , E_{i}^y , E_{i}^z ).
\end{align}
Then, the Higgsing of the rank-2 U(1) LGT is performed by ($a=x$, $y$, $z$ or $a=1$, $2$, $3$)
\begin{align}
X_{a,i} &= \exp ( i A_{i}^{a} )  , & Z_{a,i} &= \exp ( 2\pi i E_{i}^{a}/p ) . 
\label{eq:discrete2}
\end{align}
Compared to the Higgsing formula used in Eq. (\ref{eq:discrete}) for the toric codes, the roles of $X$ and $Z$ operators have been switched in Eq. (\ref{eq:discrete2}). This is by design and will be crucial in constructing the hybrid (F3) model.
The $\mathbb{Z}_p$ X-cube model~\cite{vijay16} is defined as
\begin{align}
H_{\text{XC}} = -\sum_i (\mathbb{A}_i^x+\mathbb{A}_i^y+\mathbb{A}_i^z + \mathcal{B}_i) .\label{eq:XC} 
\end{align}
The mutually commuting Hermitian projectors $\mathbb{A}_i^a$ and $\mathcal{B}_i$ are represented as
\begin{align}
\mathbb{A}_i^a &= \frac{1}{p} \sum_{j =0}^{p-1} (\mathsf{a}_i^a )^j &\mathcal{B}_i &= \frac{1}{p} \sum_{j =0}^{p-1} (b_i )^j, \label{eq:A-and-B-in-XC}
\end{align}
using a set of mutually commuting unitary operators $( \mathsf{a}_i^x , \mathsf{a}_i^y , \mathsf{a}_i^z , b_i )$ in the $p$-dimensional Hilbert space of spins constructed as
\begin{align}
\mathsf{a}_i^x  =& X_{3,i} X_{3,i-\hat{y}}^{-1} X_{2,i}^{-1} X_{2,i-\hat{z}},\nn
\mathsf{a}_i^y  =& X_{1,i} X_{1,i-\hat{z}}^{-1} X_{3,i}^{-1} X_{3,i-\hat{x}},\nn
\mathsf{a}_i^z  =& X_{2,i} X_{2,i-\hat{x}}^{-1} X_{1,i}^{-1} X_{1,i-\hat{y}},\nn
b_i =& Z_{1,i+\hat{y}+\hat{z}} Z_{1,i+\hat{y}}^{-1} Z_{1,i+\hat{z}}^{-1} Z_{1,i}\nn
\times&  Z_{2,i+\hat{x}+\hat{z}} Z_{2,i+\hat{x}}^{-1} Z_{2,i+\hat{z}}^{-1} Z_{2,i}\nn
\times& Z_{3,i+\hat{x}+\hat{y}} Z_{3,i+\hat{x}}^{-1} Z_{3,i+\hat{y}}^{-1} Z_{3,i}. \label{eq:XCeq}
\end{align}
In particular, $\mathsf{a}_i^x \mathsf{a}_i^y \mathsf{a}_i^z = 1$ following from $B_i^x + B_i^y + B_i^z = 0$ of the parent LGT.

The two types of elementary excitations in the X-cube model are called $l_a$-quasiparticles ($a=x,y,z$) and fractons. The $l_a$-quasiparticles are the excitations of the $\mathbb{A}_i^a$ operators while fractons are those of the $\mathcal{B}_i$ operator defined in Eq. (\ref{eq:A-and-B-in-XC}). The integer-valued charges of $l_a$-quasiparticles and fractons are denoted by $\rho^{l_a}$ and $\rho^f$, and are respectively determined as the eigenvalues of the operators 
\begin{align}
\mathbb{A}_i^a (n) &= \frac{1}{p} \sum_{j =0}^{p-1} (\omega^{-n} \mathsf{a}_i^a )^j  , &\mathcal{B}_i (n) &= \frac{1}{p} \sum_{j =0}^{p-1} (\omega^{-n} b_i )^j. \label{eq:ch}
\end{align}
The eigenstate $\ket{\psi_{l_a}}$ with a $l_a$ quasiparticle is defined as $\mathbb{A}_i^a (\rho^{l_a} ) |\psi_{l_a}\rangle =  |\psi_{l_a}\rangle$ and $\mathcal{B}_i (\rho^f ) |\psi_f\rangle =  |\psi_f\rangle$ defines the eigenstate with a fracton excitation. In other words, $\ket{\psi_{l_a}}$($\ket{\psi_{f}}$) is an eigenstate of $\mathsf{a}_i^a$($b_i$) with eigenvalue $\omega^{\rho^{l_a}}$($\omega^{\rho^{f}}$).

The term {\it lineon} is used to refer to the excitation in the X-cube model that can move freely only in a particular direction. They can be defined in terms of the $l_a$-quasiparticles just defined, as the eigenstates satisfying
\begin{align}
(x): & \mathbb{A}_i^x (0) |\psi_x\rangle =\mathbb{A}_i^y (-\rho^x ) |\psi_x\rangle =\mathbb{A}_i^z (\rho^x ) |\psi_x\rangle =\ket{\psi_x},  \nn
(y): & \mathbb{A}_i^x (\rho^y) |\psi_y\rangle = \mathbb{A}_i^y (0 )|\psi_y\rangle =\mathbb{A}_i^z (-\rho^y) |\psi_y\rangle =\ket{\psi_y},\nn
(z): & \mathbb{A}_i^x (-\rho^z ) |\psi_z\rangle =\mathbb{A}_i^y (\rho^z )|\psi_z\rangle=\mathbb{A}_i^z (0) |\psi_z\rangle =\ket{\psi_z}. \label{eq:lineon}
\end{align}
The $x$, $y$, $z$ lineons denoted by $(x), (y), (z)$ with charges $\rho^x , \rho^y, \rho^z$ are respectively defined as having the eigenvalues $(0,-\rho^x , \rho^x )$, $(\rho^y ,0,- \rho^y)$, and $(-\rho^z ,\rho^z ,0)$ of the $l_a$-quasiparticles. In other words, lineons are composite excitations consisting of a pair of $l_a$-quasiparticles. Hence the $l_a$-quasiparticles are the more fundamental excitations and we proceed to construct field theories for them, with the lineon dynamics emerging as a natural by-product. 

The matter field Lagrangian of fractons and $l_a$-quasiparticles can be derived as 
\begin{align}
\mathcal{L}_{f}& =i\psi_f^\dag \partial_t \psi_f- \frac{1}{2m_f} \sum_{a<b}| {\bm D}_f^{ab} \psi_f |^2 , \nn
\mathcal{L}_{l}& =i{\bm \psi}_l^\dag \cdot \partial_t {\bm \psi}_l - \frac{1}{2m_m} \sum_{a}| {\bm D}_l^{a} {\bm \psi}_l |^2  , \label{eq:Lf-and-Ll}
\end{align}
where $\psi^f$ and ${\bm \psi}_l = (\psi_l^x , \psi_l^y , \psi_l^z)$ represent the fracton and the three species of $l_a$-quasiparticle fields, respectively. The covariant derivatives are given as
\begin{align}
{\bm D}_f^{ab} \psi_f &= \psi_f \partial_a\partial_b \psi_f - \partial_a \psi_f \partial_b \psi_f+i\sum_c|\epsilon_{abc}|A^c\psi_f \nn 
{\bm D}_l^{x} {\bm \psi}_l &=\psi_l^z\partial_x \psi_l^y-\psi_l^y\partial_x \psi_l^z+iE^x\psi_l^y\psi_l^z\nn
{\bm D}_l^{y} {\bm \psi}_l &=\psi_l^z\partial_y \psi_l^x-\psi_l^x\partial_y \psi_l^z+iE^y \psi_l^x\psi_l^z\nn
{\bm D}_l^{z} {\bm \psi}_l &=\psi_l^x\partial_z \psi_l^y-\psi_l^y\partial_z \psi_l^x+iE^z\psi_l^x\psi_l^y . 
\end{align}
One can readily check their gauge covariance under $\psi_f \rightarrow e^{i f} \psi_f$, $\psi_l^a \rightarrow e^{ig^a} \psi_l^a$, as well as $A^{ab} \rightarrow A^{ab} - \partial_a \partial_b f$ and $E^{ab} \rightarrow E^{ab} + \sum_{c} \epsilon_{abc} (\partial_c g^b-\partial_c g^a)$. Various covariant derivatives given here, while straightforward to show, were not derived previously. 

The continuity equations of $l_a$-quasiparticles and fractons follow from the matter Lagrangians (\ref{eq:Lf-and-Ll}) as 
\begin{align}
&\partial_t \rho^f - \sum_{a < b} \partial_a\partial_b [j^{f}]^{ab}= 0 \nn
&\partial_t \rho^{l_a} + \sum_{bc}\epsilon_{abc} \partial_c [j^{l}]^c =0 . \label{eq:ccl1}  \end{align} 
Here $\rho^f$ and $\rho^{l_a}$ are respectively the densities of fractons and $l_a$-quasiparticles, and $[j^f]^{ab}$ and $\bm{j}^l$, respectively a symmetric tensor and a vector, represent the current density of the fracton and $l_a$-quasiparticles.  These conservation laws, first mentioned in Ref. \onlinecite{slagle17}, are rigorously derived from the matter field Lagrangian in Eq. (\ref{eq:Lf-and-Ll}). Charge and current density expressions in terms of matter fields are written in Appendix \ref{asec:D}. 

The EFT of the XC model can be constructed by incorporating the conservation laws of quasiparticles by means of Lagrange multipliers: 
\begin{align}
\mathcal{L}_{\text{XC}} & =\frac{p}{2\pi}  \sum_{abc} |\epsilon_{abc}| \frac{1}{2} A^{ab} \partial_t  E^c  -\frac{1}{2}\sum_{ab} A^{ab} [j^f]^{ab} - {\bm E} \cdot  \bm j^l \nn
& +\frac{p}{2\pi} E^{a0}  \Bigl(\sum_{bc} \epsilon_{abc} \partial_c A^{ab}- \frac{2\pi}{p} {\rho}^{l_a} \Bigl)\nn
&+A^0  \Bigl(\frac{p}{2\pi} \sum_{abc} |\epsilon_{abc}| \frac{1}{2}\partial_a\partial_b E^c-\rho^f \Bigl) .  \label{eq:XCL}
\end{align}
The sum is over spatial indices. The commutation relation and the gauge transformation of the $A^{ab}$ and $E^a$ in Eq. (\ref{eq:XCL}) are
\begin{align}
&[E^a(\bm r), A^{bc}(\bm r')]=|\epsilon_{abc}| \frac{2\pi i}{p}\delta^3(\bm r -\bm r')\nn
&A^{ab}(\bm r) \rightarrow A^{ab}(\bm r)+\partial_a\partial_b f\nn
&E^a(\bm r) \rightarrow E^a(\bm r)-\epsilon_{abc} \partial_a g^c,
\end{align}

We have given the derivation of the Lagrangian (\ref{eq:XCL}) starting from the matter Lagrangian and the subsequent conservation laws in Appendix \ref{asec:D}. The Lagrangian appears identical to the one given in Ref. \onlinecite{slagle17}, but there is a subtle difference in the way the constraints are enforced. In order to properly embed the conservation laws of $l_a$-quasiparticles and fractons as well as the identity $\sum_a \rho^{l_a}=0$ in the Lagrangian, the gauge transformations of $A^0$ and $E^{a0}$ must be 
\begin{align}
    A^0 (\bm r)&\rightarrow A^0(\bm r)+\partial_t f(\bm r)\nn
    \hspace{0cm}E^{a0} (\bm r) &\rightarrow E^{a0} (\bm r) + \partial_t g^a (\bm r) + g^0(\bm r) .
\end{align}
Conservation laws of $l_a$-quasiparticles and fractons follow from requiring gauge invariance of the Lagrangian. In particular, the gauge transformation involving $g^0$ gives the term $g^0 \sum_a \rho^{l_a}$, which leads to the constraint $\sum_a \rho^{l_a}=0$.~\footnote{A different method was used in Ref. \cite{slagle17} to enforce this constraint.}

The lattice model for fracton fields $\psi^f_i$ can be constructed from the gauge principle. Given that fracton fields couple to ${\bm A}$ which transforms as $A_i^{ab} \rightarrow A^{ab}_i - f_{i-\hat{a}-\hat{b}}+f_{i-\hat{a}}+f_{i-\hat{b}}-f_{i}$ [Eq. (\ref{eq:gt2})], we write down the gauge-invariant hopping of the fractons as
\begin{align}
(\psi^f )^\dag_{i-\hat{b}} (\psi^f )^\dag_{i-\hat{a}} \psi^f_{i-\hat{a}-\hat{b}} \psi^f_i e^{i A^{ab}_i } \label{eq:flm}
\end{align}
and its Hermitian conjugate. The fracton motion clearly preserves the center-of-mass, or the dipole moment. Note that this is a three-dimensional extension of the quadrupole model studied in the context of quadrupole topological insulators~\cite{hughes20, hughes21b}. 

The $l_a$-quasiparticles coupled to the $E^{ab}$ fields which transform as in Eq. (\ref{eq:gt2}) have the gauge-invariant hopping terms
\begin{align}
&(\psi^{l_x})_{i}^\dag (\psi^{l_y})_{i+\hat{z}}^\dag \psi^{l_x}_{i+\hat{z}} \psi^{l_y}_{i}  e^{iE^{xy}_i},\nn
&(\psi^{l_y})_{i}^\dag (\psi^{l_z})_{i+\hat{x}}^\dag \psi^{l_y}_{i+\hat{x}} \psi^{l_z}_{i}  e^{iE^{yz}_i},\nn
&(\psi^{l_z})_{i}^\dag (\psi^{l_x})_{i+\hat{y}}^\dag \psi^{l_z}_{i+\hat{y}} \psi^{l_x}_{i}  e^{iE^{zx}_i}
\end{align}
and their Hermitian conjugates. The first line, for instance, clearly shows that a $z$-lineon motion is made by the coordinated motion of the $l_x$ and $l_y$ quasiparticles in the opposite directions along the $z$-axis. 

The lattice formulation of the fracton and lineon dynamics coupled to the rank-2 gauge fields was not discussed previously, while their derivation is quite straightforward from the gauge principle as shown above. 

\section{Hybrid rank-1 and rank-2 LGT and F3 model as its descendant}
\label{sec:3}

We finally come to the discussion of the F3 model that we proposed in an earlier publication. We first lay out the underlying LGT of the F3 model and its Higgsing, and then construct the relevant effective field theory. 

\subsection{F3 model from Higgsing the hybrid rank-1 and rank-2 LGT}
There are two types of terms in the X-cube model, and we ask if any of them can be added onto the 3DTC without violating the exact solvability. One sees readily that the addition of the magnetic field terms $\mathsf{a}_i^a$ from the X-cube model [Eq. (\ref{eq:XCeq})] to the 3DTC poses a problem because they in general fail to commute with the magnetic flux operators $c_i^a$ in the 3DTC: $[\mathsf{a}_i^a,c_j^{a'}]\neq 0$. 

The Gauss's law operator ${\cal B}_i$ of the X-cube model obviously commutes with the magnetic fields ${\cal C}_i^a$ of the 3DTC as they are built out of the same $Z$ operators. On the other hand, the ${\cal B}_i$'s commute with the Gauss's law operator ${\cal A}_i$ of the 3DTC {\it only if} the local Hilbert space dimension is $p=3$ since commuting the two operators results in the phase $\omega^3$, which equals one only for $p=3$ or its multiple. 
This is the {\it compatibility condition} between the Gauss's laws of the rank-1 and rank-2 U(1) LGTs. When the compatibility condition is fulfilled, we obtain a new, exactly solvable spin Hamiltonian 
\begin{align}
H_{\text{F3}}&=-\sum_i \mathcal{A}_i-\sum_i \mathcal{B}_i- \sum_{i,a} \mathcal{C}_{i}^a,    
\end{align}
which is the F3 model~\cite{kim21}. This model is the sum of the 3DTC ($\mathcal{A}_i$ and $\mathcal{C}_i^a$)  and the Gauss's law $\mathcal{B}_i$ of the X-cube model. 
The quasiparticle excitations associated with $\mathcal{A}_i , \mathcal{B}_i$ and $\mathcal{C}^a_i$ were respectively dubbed the freeon, fracton, and the fluxon, hence the F3 model. One can think of F3 as the 3DTC modified by the presence of the cube term $-\sum_i \mathcal{B}_i$. Despite the additional term, the logical operators of the F3 model are identical to those of the 3DTC and consequently they share the same GSD. This observation will permeate the field-theoretic discussion that follows as well. 

We can derive the F3 model by Higgsing the gauge theory of two vector gauge fields $\bm A_i$ and $\bm E_i$ which satisfy the commutation relation $[A_i^a,E_j^b]=i\delta_{ij} \delta_{ab}$. 
One can identify the three ``mutually commuting" generators, $(\bm \nabla \cdot \bm A)_i$, $(\bm \nabla \times \bm E)_i$, and $({\bf DE})_i$, given by 
\begin{align}
({\bm \nabla} \cdot {\bm A})_i &\equiv \sum_{a}  (A_i^a-A_{i-\hat{a}}^a)\nn
(\bm \nabla \times \bm E)_i &\equiv  \sum_{b,c} \epsilon_{abc}( E_i^b-E_{i+\hat{c}}^b)\nn
({\bf DE})_{i} &\equiv E_{i +\hat{x}+\hat{y}}^{z} - E_{i+\hat{x}}^{z} - E_{i+\hat{y}}^{z} + E_{i}^{z}\nn
&~~~~+E_{i+\hat{y}+\hat{z}}^{x}-E_{i+\hat{y}}^{x}-E_{i+\hat{z}}^{x}+E_{i}^{x}\nn
&~~~~+E_{i+\hat{x}+\hat{z}}^{y}-E_{i+\hat{x}}^{y}-E_{i+\hat{z}}^{y}+E_{i}^{y} . \label{eq:F3def}
\end{align}
One can recognize the first two generators as those of the rank-1 LGT in Eq. (\ref{eq:div}), except for the reversal of the roles ${\bm A} \leftrightarrow {\bm E}$. The rank-2 U(1) generator $({\bf D}{\bf E})_i$ is taken from Eq. (\ref{eq:R2-Gauss-law}). In fact, we do not have strict commutativity of all three operators since the explicit calculation shows 
\begin{align}
[({\bm \nabla} \cdot {\bm A})_i, ({\bf DE})_{j}]&=3i(\delta_{ij}+\delta_{i,j+\hat{x}+\hat{y}}+\delta_{i,j+\hat{x}+\hat{z}}\nn
&~~~+\delta_{i,j+\hat{y}+\hat{z}}-\delta_{i,j+\hat{x}}-\delta_{i,j+\hat{y}}\nn
&~~~-\delta_{i,j+\hat{z}}-\delta_{i,j+\hat{x}+\hat{y}+\hat{z}}) . \label{eq:A-and-E-commute} 
\end{align}
Nevertheless, we can show that the commutativity is recovered after the Higgsing, thus justifying the use of ``mutually commuting" operators in parenthesis. 

The Higgsing formula is given by
\begin{align}
X_{a,i} &= \exp ( i A_{i}^{a} )  , & Z_{a,i} &= \exp ( 2\pi i E_{i}^{a}/p ).
\end{align}
One can easily identify
\begin{align} \exp((\bm \nabla \cdot \bm A)_i) & \equiv a_i \nn 
\exp ( 2\pi i ({\bf DE} )_i /p  ) & \equiv b_i \nn 
\exp(2\pi i(\bm \nabla \times \bm E)^a_i/p) & \equiv c_i^a 
\end{align} 
as the $a_i,~ b_i, ~ c_i^a$ operators of the F3 model. By virtue of Eq. (\ref{eq:A-and-E-commute}), we see that the commutator $[({\bm \nabla} \cdot {\bm A})_i, ({\bf DE})_{j}]$ equals $+3i, -3i$ or 0, and therefore $[a_i, b_j ] =0$ provided $p=3$. That is, although the generators themselves do not commute, their Higgsed versions do commute in the special case of $p=3$. 

%
%

The gauge transformation rules of $\bm A_i$ and $\bm E_i$ are 
\begin{align}
A^a_i &\rightarrow U_A^\dag  A_i^a U_A \nn
A^x_i &\rightarrow A^x_i- f^0_{i-\hat{y}-\hat{z}}+f^0_{i-\hat{y}}+f^0_{i-\hat{z}}-f^0_{i}\nn
&~~~-f_i^z+f_{i-\hat{y}}^z+f_i^y-f_{i-\hat{z}}^y\nn
&\sim A_i^x-\partial_y\partial_z f_i^0-\partial_y f_i^z+\partial_z f_i^y\nn
A^y_i &\rightarrow A^y_i- f^0_{i-\hat{x}-\hat{z}} +f^0_{i-\hat{x}} +f^0_{i-\hat{z}}-f^0_{i}\nn
&~~~-f_i^x+f_{i-\hat{z}}^x+f_i^z-f_{i-\hat{x}}^z\nn
&\sim A_i^y-\partial_z\partial_x f_i^0-\partial_z f_i^x+\partial_x f_i^z\nn
A^z_i &\rightarrow A^z_i -  f^0_{i-\hat{x}-\hat{y}} + f^0_{i-\hat{x}} + f^0_{i-\hat{y}} - f^0_{i}\nn 
&~~~-f_i^y+f_{i-\hat{x}}^y+f_i^x-f_{i-\hat{y}}^x\nn
&\sim A_i^z-\partial_x\partial_y f_i^0-\partial_x f_i^y+\partial_y f_i^x\nn
E_i^a &\rightarrow U_E^\dag E_i^a U_E =E_i^a- g_{i+\hat{a}}+g_i\nn
&\sim E_i^a -\partial_a g_i.
\label{eq:A-E-F3} \end{align}
The two unitary operators are 
\begin{align}
U_A & = \exp \left( i \sum_j (f^0_j ({\bf DE})_j +{\bm f}_j \cdot (\bm \nabla \times \bm E)_j ) \right) \nn 
U_E & = \exp \left( i \sum_j g_j ({\bm \nabla} \cdot {\bm A})_j \right), \label{eq:Ua-and-Ue-F3}
\end{align}
for $\bm f=(f^x,f^y,f^z)$. (When $f^0_j = 0$, these unitary operators are those of the 3DTC with the roles of $\bm A$ and $\bm E$ interchanged.) Since the generators do not commute, $[({\bm \nabla} \cdot {\bm A})_i, ({\bf DE})_{j}] \neq 0$, neither do the unitary operators $U_A U_E \neq U_E U_A$. An explicit calculation shows
\begin{widetext}
\begin{align}
    [ \sum_j f^0_j ({\bf DE})_j , \sum_i g_i ({\bm \nabla} \cdot {\bm A})_i ] & = -3i \sum_i g_i [ f^0_i + f^0_{i-\hat{x}-\hat{y}} + f^0_{i-\hat{y}-\hat{z}} + f^0_{i-\hat{z}-\hat{x}}  - f^0_{i-\hat{x}} - f^0_{i-\hat{y}} - f^0_{i-\hat{z}} - f^0_{i-\hat{x}-\hat{y}-\hat{z}} ] \nn 
    & = -3i \sum_i f^0_i [g_i + g_{i+\hat{x}+\hat{y}} + g_{i+\hat{y}+\hat{z}} + g_{i+\hat{z}+\hat{x}}  - g_{i+\hat{x}} - g_{i+\hat{y}} - g_{i+\hat{z}} - g_{i+\hat{x}+\hat{y}+\hat{z}} ]  . 
\end{align}
\end{widetext}
In the continuum limit one can show that the commutator reduces to $-3i \sum_i g_i \partial_x \partial_y \partial_z f^0_i = +3i \sum_i f^0_i \partial_x \partial_y \partial_z g_i$. By choosing $f^0$ or $g$ to be functions of a special type, i.e. 
\begin{align}
    f^0_i = f^0_1 (i_x , i_y ) + f^0_2 (i_y , i_z ) + f^0_3 (i_z , i_x ) 
\end{align}
(or similarly for $g_i$), one can prove the commutator vanishes $[ \sum_j f^0_j ({\bf DE})_j , \sum_i g_i ({\bm \nabla} \cdot {\bm A})_i ] = 0$. We thus obtain valid definitions of unitary operators $U_A$ and $U_E$ with $[U_A , U_E ] =0$ under the (mild) restriction on the gauge functions $f^0_i$ or $g_i$. Such a restriction is lifted when we turn off $f^0_i =0$ and the model becomes identical to 3DTC. These transformation rules (and the constraint on the gauge functions) will play a vital role in identifying the field theory description of the F3 model. 

\subsection{Field theory of F3 model}

The F3 model was derived from the hybrid of the rank-1 and rank-2 U(1) LGTs. We now apply the gauge principle to derive the appropriate EFT of the F3 model. 
There are three types of quasiparticle excitations in the F3 model called the freeons, fractons, and fluxons, which are the excitations of the $\mathcal{A}_i$, $\mathcal{B}_i$, and $\mathcal{C}^a_i$ operators. We will use $e$, $f$, $m$ to express the quantities related to freeons, fractons, fluxons, respectively.  

First of all, the gauge transformation of $\bm A$ and $\bm E$ in the continuum field theory follows directly from the lattice transformation rules given in Eq. (\ref{eq:A-E-F3})
\begin{align}
A^a ({\bm r})&\rightarrow U_A^\dag  A^a ({\bm r}) U_A \nn
& = A^a ({\bm r}) - \frac{1}{2} \sum_{bc} |\epsilon_{abc} | \partial_b \partial_c f^0 ({\bm r}) - ({\bm \nabla} \times \bm f({\bm r}))^a \nn 
%
\bm E ({\bm r})&\rightarrow U_E^\dag \bm E ({\bm r}) U_E =\bm E ({\bm r})-{\bm \nabla} g ({\bm r}). \label{eq:3D-transform2} 
\end{align}
The two unitary operators
\begin{align}
U_A & = \exp \left[ i\int d^3 {\bm r} \Bigl(f^0 ({\bm r} ) {\bf DE}(\bm r ) +{\bm f}(\bm r) \cdot \bm \nabla \times \bm E(\bm r ) \Bigl)\right]\nn 
U_E & = \exp \left[i \int d^3 {\bm r} g({\bm r} ) {\bm \nabla} \cdot {\bm A} ({\bm r}) \right],
\end{align}
$\bm f=(f^x,f^y,f^z)$, are the continuum descendants of the lattice operators given in Eq. (\ref{eq:Ua-and-Ue-F3}). In particular, we have ${\bf DE} = \partial_x \partial_y E^z + \partial_y \partial_z E^x + \partial_x \partial_y E^z$. As in the lattice consideration, the two unitary operators commute provided we choose either $\partial_x \partial_y \partial_z f^0 ({\bm r}) = 0$ or $\partial_x \partial_y \partial_z g ({\bm r}) =0$. Such condition is fulfilled by choosing $f^0$ or $g$ to be the sum of functions that depend on at most two of the coordinates but not on all three. 

The freeon field $\psi_{e}$ transforms as $U_E^\dag \psi_{e} U_E = e^{ig} \psi_{e}$ under the constraint 
\begin{align} {\bm \nabla} \cdot {\bm A} = \rho_{e} = \psi^\dag_{e} \psi_{e}.  \end{align} 
It behaves in the same way as the electric quasiparticle in the 3DTC with the matter field Lagrangin in Eq. (\ref{eq:TCL}) and the continuity equation $\partial_t \rho^e +\nabla \cdot \bm j^e=0$. The other quasiparticles, i.e. fracton and three species of fluxons, can be grouped as $\psi^\mu$ ($\mu =0$ for fracton, $\mu=x,y,z$ for fluxons) and transform as 
\begin{align}
\psi^\mu\rightarrow U_A^\dag \psi^\mu U_A = e^{i f^\mu}\psi^\mu 
\end{align} 
under the constraints
\begin{align}
{\bf DE}(\bm r )  = \rho^f  , ~~ {\bm \nabla} \times {\bm E} = {\bm \rho}^m   \label{eq:F3-constraints}
\end{align}
where ${\bm \rho}^m = (\rho^{m_x} , \rho^{m_y} , \rho^{m_z} )$.
A gauge-invariant Lagrangian can be constructed as 
\begin{align}
\mathcal{L}_f=i\sum_\mu (\psi^\mu )^\dag \partial_t \psi^\mu-\frac{1}{2m} \sum_{a < b} |\bm D^{ab} \psi|^2 \label{eq:Lf2}
\end{align}
with the covariant derivatives given by
\begin{align}
\bm D^{xy} \psi&= \psi^x \psi^y \psi^0 \partial_x \partial_y \psi^0 - \psi^x \psi^y \partial_x \psi^0 \partial_y \psi^0\nn
&+{(\psi^0)}^2 [ \psi^x \partial_x \psi^y- \psi^y \partial_y \psi^x+i A^z \psi^x \psi^y ] \nn
\bm D^{xz} \psi&= \psi^x \psi^z \psi^0 \partial_x \partial_z \psi^0-\psi^x \psi^z \partial_x \psi^0 \partial_z \psi^0\nn
&+{(\psi^0)}^2 [ \psi^z \partial_z \psi^x- \psi_f^x \partial_x \psi^z+i A^y \psi^y\psi^z ] \nn
\bm D^{yz} \psi &= \psi^y \psi^z \psi^0 \partial_y \partial_z \psi^0-\psi^y \psi^z \partial_y \psi^0 \partial_z \psi^0\nn
&+{(\psi^0)}^2 [ \psi^y \partial_y \psi^z- \psi^z \partial_z \psi^y+i A^x \psi^y\psi^z ] .
\end{align}
They are quartic in the matter fields and involve the interaction between fractons and fluxons. (The Lagrangian itself is octic in the matter fields.) Once the fracton field is turned off by taking $\psi^0 = 1$, we recover the covariant derivative of the 3DTC given in Eq. (\ref{eq:3DTC-cov-der}), which again reveals the 3DTC root of the F3 model. 

The conservation laws following from the Lagrangian in Eq. (\ref{eq:Lf2}) are
\begin{align}
\partial_t \rho^f &  =  \partial_x \partial_y j^z + \partial_x \partial_z j^y + \partial_y \partial_z j^x , \nn
\partial_t \bm \rho^{m}&  =  \nabla \times \bm j \label{eq:ccl3} .  \end{align} 
The quasiparticle density and the current operators are given by
\begin{align}
\rho^{f} & = {(\psi^0)}^\dag \psi^0  \nn
{\bm \rho}^{m} & = ( (\psi^x)^\dag \psi^x , (\psi^y )^\dag \psi^y , (\psi^z )^\dag \psi^z ) \nn
j^x &= \frac{1}{2m i} \left[ {(\psi^y)}^\dag {(\psi^z)}^\dag {(\psi^0)}^\dag {(\psi^0)}^\dag {\bm D}^{yz}\psi - h.c. \right]\nn
j^y &= \frac{1}{2m i} \left[ {(\psi^z)}^\dag {(\psi^x)}^\dag {(\psi^0)}^\dag {(\psi^0)}^\dag {\bm D}^{xz}\psi - h.c. \right]\nn
j^z &= \frac{1}{2m i} \left[ {(\psi^x)}^\dag {(\psi^y)}^\dag {(\psi^0)}^\dag {(\psi^0)}^\dag {\bm D}^{xy}\psi - h.c. \right]. \label{eq:cle2}
\end{align}
First of all, note that $\partial_t {\bm \rho}^m = {\bm \nabla} \times {\bm j}$ is identical to the conservation of magnetic charges in Eq. ({\ref{eq:3DTC-curr-cons-simplified}}). 
Furthermore, the constraint ${\bm \rho}^m = {\bm \nabla} \times {\bm E}$ given in Eq. (\ref{eq:F3-constraints}) implies $\bm \nabla \cdot \bm \rho^m=0$, i.e. the magnetic excitations must form closed loops. Time derivatives of both fracton ($\rho^f$) and the fluxon ($\bm \rho^m$) densities are related to the same current density vector $\bm j$ in Eq. (\ref{eq:ccl3}), which goes to show that fractons and fluxons are not independent entities in the F3 model, but are intertwined objects through the definitions of the current operators ${\bm j}$ given above and the continuity equation given in Eq. (\ref{eq:ccl3}). 

Having established the EFT of the freeon, fracton, and fluxon matter fields, we proceed to construct the gauge field Lagrangian. The starting point, as usual, is the constraints relating the matter and the gauge degrees of freedom:
\begin{align}
(\bm \nabla \cdot \bm A)_i &= \frac{2\pi}{p}\rho^e_{i}, & (\bm \nabla \times \bm E)_i &= \bm \rho^m_i, & ({\bf DE})_i &= \rho^f_{i}.\label{eq:cf31}
\end{align}
Taking the time derivatives on both sides of the equations and invoking the continuity equation of freeons in Eq. (\ref{eq:3dtccl}), and of fluxons and fractons in Eq. (\ref{eq:ccl3}), we obtain three relations 
\begin{align}
(\bm \nabla \cdot  \partial_t {\bm A})_i&=-(2\pi/p)(\nabla \cdot \bm j^e)_i\nn
(\bm \nabla \times  \partial_t {\bm E})_i&=(\nabla \times \bm j)_i\nn
\partial_t ({\bf DE})_i&=(\partial_x\partial_y j^z+\partial_y\partial_z j^x+\partial_x\partial_z j^y)_i\label{eq:cf32}
\end{align}
that are solved by 
\begin{align}
\partial_t {{\bm A}}_i&=-\frac{2\pi}{p} {\bm j}^e_i & \partial_t {\bm E}_i &={\bm j}_i. \label{eq:fa}
\end{align}
Luckily the second and the third equations in Eq. (\ref{eq:cf32}) share the same solution, providing consistency to the whole approach we take. The Lagrangian that embodies both Eqs. (\ref{eq:cf31}) and (\ref{eq:fa}) is readily constructed:
\begin{align}
\mathcal{L}_{\rm F3} & =  {\bm E}_i \cdot \partial_t {\bm A}_i +\bm A_i \cdot \bm j_i + \frac{2\pi}{p} {\bm E}_i \cdot  \bm j^e_i \nn
& +\frac{p}{2\pi} E^{0}_i  \Bigl( \bm \nabla \cdot \bm A - \frac{2\pi}{p} {\rho}^e \Bigl)_{i} + \bm A^0_i \cdot  \Bigl(\bm \nabla \times \bm E-{\bm \rho}^m \Bigl)_i \nn 
&+A^0_i  \Bigl( {\bf DE} - \rho^f \Bigl)_i.
\end{align}
Several Lagrange multipliers, $(E^0 , A^0 , {\bm A}^0 )$, are introduced to implement the Gauss constraints relating the gauge fields to the matter fields. 
By redefining $\bm A \rightarrow -\bm A$ and $-(2\pi/p) \bm E\rightarrow \bm E$, one can rewrite the F3 Lagrangian as
\begin{align}
\mathcal{L}_{\text{F3}} & = \frac{p}{2\pi}  {\bm E} \cdot \partial_t {\bm A} -\bm A \cdot \bm j - {\bm E} \cdot  \bm j^e \nn
&- \frac{p}{2\pi} E^{0} \Bigl(\bm \nabla \cdot \bm A + \frac{2\pi}{p} {\rho}^e \Bigl) - \bm A^0 \cdot  \Bigl(\frac{p}{2\pi}\bm \nabla \times \bm E +  {\bm \rho}^m \Bigl)\nn
&- A^0  \Bigl(\frac{p}{2\pi}{\bf DE} + \rho^f \Bigl)\nn
& = \mathcal{L}_{\rm BF} - A^0  \Bigl( \frac{p}{2\pi}{\bf DE} + \rho^f \Bigl) , \label{eq:Lf3}
\end{align}
clearly showing $\mathcal{L}_{\rm F3}$ as the action of the 3D toric code $\mathcal{L}_{\rm BF}$ given in Eq. (\ref{eq:EFT-for-3DTC}), supplemented with an additional constraint through the Lagrange multiplier $A^0$. The constraint in turn ties the fracton density $\rho^f$ to the local configuration of the ${\bf E}$ through the relation ${\bf DE} + \rho^f =0$. Note that there is no other term in the effective action that depends on the fracton degrees of freedom. The current ${\bf j}$ is also defined as a fracton-fluxon composite object as seen in Eq. (\ref{eq:cle2}). All these features are indications that fracton dynamics is `parasitic' to that of fluxons, and the overall dynamics resembles that of 3D toric code. 

The commutation relation and the gauge transformation of the $\bm A$ and $\bm E$ in Eq. (\ref{eq:Lf3}) are
\begin{align}
&[A^a(\bm r), E^{b}(\bm r')]=\frac{2\pi i}{p}\delta_{ab} \delta^3(\bm r -\bm r')\nn
&A^a ({\bm r})\rightarrow U_A^\dag  A^a ({\bm r}) U_A\nn
&~~~~~~~~= A^a ({\bm r}) + \frac{1}{2} \sum_{bc} |\epsilon_{abc} | \partial_b \partial_c f^0 ({\bm r}) + ({\bm \nabla} \times \bm f({\bm r}))^a\nn
&\bm E(\bm r) \rightarrow U_A^\dag  A^a ({\bm r}) U_A = \bm E ({\bm r})+{\bm \nabla} g ({\bm r})
\label{eq:F3-space-transform} 
\end{align}
where $U_A$ and $U_E$ are
\begin{align}
U_A &= \exp \left[ -\frac{ip}{2\pi}\int d^3 {\bm r} \Bigl(f^0 ({\bm r} ) {\bf DE}(\bm r ) +{\bm f}(\bm r) \cdot \bm \nabla \times \bm E(\bm r ) \Bigl)\right]\nn
U_E &= \exp \left[-\frac{ip}{2\pi} \int d^3 {\bm r} g({\bm r} ) {\bm \nabla} \cdot {\bm A} ({\bm r}) \right].
\end{align}
The gauge transformations of $\bm A^0$,  $A^0$, and $E^0$ are 
\begin{align}
\bm A^0(\bm r) &\rightarrow \bm A^0(\bm r)+\partial_t \bm f(\bm r)+\bm \nabla f'(\bm r)\nn
A^0(\bm r) &\rightarrow A^0(\bm r)+\partial_t f^0(\bm r)\nn
E^{0}(\bm r) &\rightarrow E^{0}(\bm r) + \partial_t g^a(\bm r). \label{eq:F3-0-transform}
\end{align}
Conservation laws of quasiparticles in Eq. (\ref{eq:ccl3}) as well as the identity $\bm \nabla \cdot \bm \rho^m=0$ follow from requiring the gauge invariance of the Lagrangian in Eq. (\ref{eq:Lf3}) under the gauge transformations in Eqs. (\ref{eq:F3-space-transform}) and (\ref{eq:F3-0-transform}).
This completes the derivation of the EFT for the F3 model.


We only need to discuss the lattice model for fracton-fluxon matter fields $\psi^{\mu}_i$ because the lattice model for the freeon field is simply the expression in Eq. (\ref{eq:lelec}). Considering the transformation of ${\bm A}$ field given in Eq. (\ref{eq:A-E-F3}), the appropriate gauge-invariant hoppings are
\begin{align}
(\psi^0 )^\dag_{i-\hat{y}} (\psi^0 )^\dag_{i-\hat{z}} \psi^0_{i-\hat{y}-\hat{z}} \psi^0_{i}  (\psi^z )^\dag_{i-\hat{y}} (\psi^y )^\dag_{i} \psi^z_{i} \psi^y_{i-\hat{z}} e^{i A^{x}_i }\nn
(\psi^0 )^\dag_{i-\hat{z}} (\psi^0 )^\dag_{i-\hat{x}} \psi^0_{i-\hat{z}-\hat{x}} \psi^0_{i} (\psi^x )^\dag_{i-\hat{z}} (\psi^z )^\dag_{i} \psi^x_{i} \psi^z_{i-\hat{x}} e^{i A^{y}_i }\nn
(\psi^0 )^\dag_{i-\hat{x}} (\psi^0 )^\dag_{i-\hat{y}} \psi^0_{i-\hat{x}-\hat{y}} \psi^0_{i}  (\psi^y )^\dag_{i-\hat{x}} (\psi^x )^\dag_{i} \psi^y_{i} \psi^x_{i-\hat{y}} e^{i A^{z}_i }
\end{align}
and their Hermitian conjugate. By comparing to Eq. (\ref{eq:2.10}) and Eq. (\ref{eq:flm}), one can see that the fracton part $\psi^0$ hops in the manner of a fracton in the X-cube model, while $\psi^a$ ($a=x,y,z$) hop as the three magnetic charges in the 3D toric code. 

As seen in Eq. (\ref{eq:F3def}), the same ${\bf E}$ field is used to define both the fluxon and the fracton charges, but these two charges are tightly bound and move as composite objects, as the above tight-binding model clearly shows. This, as well as the effective field theory of the F3 model given as the sum of the BF model plus a constraint as in Eq. (\ref{eq:Lf3}) indicates that both F3 and the toric code share the same universal physics.

\section{Summary} \label{sec:5}
We have developed an easy-to-implement scheme for deriving field theories of toric code and X-cube model in three dimensions, and applied the method to derive the field theory for the F3 model that two of the present authors wrote down earlier~\cite{kim21}. We show that the F3 model follows from Higgsing the hybrid of rank-1 and rank-2 U(1) lattice gauge theories in three dimensions. By consistently applying the gauge principle, we are able to derive the matter and the gauge field Lagrangian for the F3 model in the continuum and also in the lattice form. It turns out the effective field theory of the F3 model is that of the 3D toric code modified by a term, and describe the same universal physics. As a by-product, we construct the matter field Lagrangians both in the continuum and in the lattice form for the well-known 3D toric code and the X-cube model as well.

\acknowledgments H. J. H. was supported by the Quantum Computing Development Program (No. 2019M3E4A1080227). He also acknowledges financial support from EPIQS Moore theory centers at Harvard and MIT. J. K. was supported by the education and training program of the
Quantum Information Research Support Center, funded through the National
research foundation of Korea(NRF) by the Ministry of science and ICT(MSIT) of
the Korean government(No.2021M3H3A103657313). Additional support came from National Research Foundation of Korea under Grant NRF-2021M3E4A1038308 and NRF-2021M3H3A1038085. 
Y.-T. O. was supported by National Research Foundation of Korea under Grant NRF-2021R1A2C4001847, NRF-2020R1A4A3079707, and NRF-2022R1I1A1A01065149. 

\appendix

\section{Derivation of the field theory of three-dimensional toric code}
\label{asec:B}
From the matter action of $e$ quasiparticles in Eq. (\ref{eq:TCL}), the conservation law for the $e$ quasiparticle is, as expected, $\partial_t \rho^e + {\bm \nabla} \cdot {\bm j}^e =  0$ with $\rho^{e}  = (\psi_e)^\dag \psi_e$ and ${\bm j}^e = [\psi_e^\dag {\bm D}_e \psi_e - h.c. ]/(2m_e i)$. For the magnetic quasiparticle part, we obtain 
\begin{eqnarray}
\partial_t \rho^{m_a} + \sum_{b} \partial_b [j^m]^{ab} & = & 0, 
\label{eq:3DTC-cons} \end{eqnarray} 
where
\begin{align}
\rho^{m_a} & = (\psi_m^a)^\dag \psi_m^a  ,\nn
{[j^m]}^{ab} &= \frac{1}{2m_m i}[(\psi_m^a)^\dag(\psi_m^b)^\dag {\bm D}_m^{ab} \psi_m^a \psi_m^b-h.c.]. \label{eq:jmab}
\end{align}
%
%

The gauge part of the EFT is constructed by first examining the various constraints on the gauge fields and the matter fields. We derive the necessary constraints directly from the Higgsing formulas in Eq. (\ref{eq:toricab}). One can read off 
\begin{align}
( {\bm \nabla } \times {\bm A} )_{i_d} & = \frac{2\pi}{p} ( {\bm \rho}^m )_{i_d} , \nn 
( {\bm \nabla } \cdot {\bm E} )_{i} & = -\rho^e_{i},
\label{eq:flux-attach-3} 
\end{align}
relating the lattice curl of $\bm A$ to the vector charge density $( {\bm \rho}^m )_{i_d} = (\rho^{m_x} , \rho^{m_y} , \rho^{m_z} )_{i_d}$ and the lattice divergence ${\bm \nabla} \cdot {\bm E}$ to the scalar charge density $\rho^e$. Taking the time derivatives on both sides of the equations and invoking the quasiparticle conservation laws in Eq. (\ref{eq:3DTC-cons}), we get $( {\bm \nabla } \times \partial_t {\bm A} )_{i_d}^a = -(2\pi /p ) (\sum_b \partial_b [j^m ]^{ab})_{i_d}$ and $( {\bm \nabla } \cdot \partial_t {\bm E} )_{i} = ({ \bm \nabla } \cdot {\bm j}^e )_{i}$ whose solutions are 
\begin{align}
\partial_t {A}^a_{i}  & = - \frac{\pi}{p} \sum_{bc} \epsilon_{abc} [j^m]_{i_d}^{bc} , \nn 
\partial_t {\bm E}_i  & =  {\bm j}^e_i  . \label{eq:flux-attach-4b} 
\end{align}
The equations (\ref{eq:flux-attach-3}) and (\ref{eq:flux-attach-4b}) formalize the constraints that are inherent in the 3DTC and can be encoded in the Lagrangian
\begin{align}
\mathcal{L} & =  {\bm E}_i \cdot \partial_t {\bm A}_i  + {\bm A}_i \cdot {\bm j}^e_i + \frac{\pi}{p} \sum_{abc} \epsilon_{abc} E^a_i [j^m]_{i_d}^{bc} \nn
& +\frac{p}{2\pi} {\bm E}^{0}_{i} \cdot \Bigl( {\bm \nabla } \times {\bm A} - \frac{2\pi}{p} {\bm \rho}^{m} \Bigl)_{i_d} \nn
&-A^0_i  \Bigl( {\bm \nabla } \cdot {\bm E} +  \rho^e  \Bigl)_i .
\end{align}
Here, ${\bm E}^0_{i} = ( E^{x0}_{i} , E^{y0}_{i} , E^{z0}_{i} )$ and $A^0_i$ are the Lagrange multipliers. Carrying out the re-definition
\begin{align}
-(2\pi/p) \sum_a \epsilon_{abc} E^a_i & \rightarrow E^{bc}_i\nn
\bm A_i \rightarrow -\bm A_i,
\end{align}
and taking the continuum limit, we can get the BF action in Eq. (\ref{eq:EFT-for-3DTC}). 
%
$A$ and $j^{e}$ are
\begin{align}
A &= (A^0, {\bm A} ) &j^e &= (\rho^e , {\bm j}^e ),
\end{align}
and $E^{\mu\nu}$ and ${[j^m]}^{\mu\nu}$ are
\begin{align}
E^{\mu \nu} &=\begin{pmatrix}
  0 & -E^{x0} & -E^{y0} & -E^{z0}\\ 
  E^{x0} &  &  & \\
  E^{y0} &  & E^{ab} & \\
  E^{z0} &  &  & 
  \end{pmatrix}\nn
{[j^m]}^{\mu \nu} &=\begin{pmatrix}
  0 & -\rho^{m_x} & -\rho^{m_y} & -\rho^{m_z}\\ 
  \rho^{m_x} &  &  & \\
  \rho^{m_y} &  & [j^m]^{ab} & \\
  \rho^{m_z} &  &  & 
  \end{pmatrix} . 
\end{align}

\section{Derivation of the field theory of X-cube model}
\label{asec:D}

The conservation laws of fracton and $l_a$-quasiparticles follow from the effective Lagrangians (\ref{eq:Lf-and-Ll}) as 
\begin{align}
&\partial_t \rho^f - \sum_{a < b} \partial_a\partial_b [j^{f}]^{ab}= 0 , \nn
&\partial_t \rho^{l_a} + \sum_{bc}\epsilon_{abc} \partial_c [j^{l}]^c =0. \end{align} 
The quasiparticle density and the current operators are given by
\begin{align}
\rho^{f} & = \psi_f^\dag \psi_f  ,\nn
\hspace{0cm} [j^{f}]^{ab} &= \frac{1}{2m_f i}\left[ \psi_f^\dag \psi_f^\dag {\bm D}_f^{ab} \psi_f- h.c. \right], \nn
\rho^{l_a} & = (\psi_l^a)^\dag \psi_l^a ,\nn
\hspace{0cm}[j^{l}]^{a} &= -\frac{1}{4m_l i} \left[ \sum_{bc}|\epsilon_{abc}| (\psi_l^b)^\dag (\psi_l^c)^\dag{\bm D}_l^a \bm \psi_l - h.c. \right]. \label{eq:cle}
\end{align}
The fracton current density tensor is symmetric, $[j^f]^{ab}=[j^f]^{ba}$.
Similar to the derivation of EFT of 3DTC, we identify the constraints directly from the Higgsing formula in Eq. (\ref{eq:XCeq}). The first of the constraints is
\begin{align}
\frac{1}{2} \sum_{abc}|\epsilon_{abc}| (\partial_a \partial_b E^c)_i=\rho^f_{i}\label{eq:flux-attach-5}.
\end{align}
Taking the time derivative on both sides and invoking the fracton conservation given by Eq. (\ref{eq:ccl1}), we get $(1/2)\sum_{abc}|\epsilon_{abc}| (\partial_a \partial_b \dot{E}^c)_i= \sum_{a < b} (\partial_a \partial_b {[j^f]}^{ab})_i$ and its solution 
\begin{align}
\dot{E}^a_i=\frac{1}{2}\sum_{bc}|\epsilon_{abc}|{[j^f]}^{bc}_i.
\end{align}
The second constraint is 
\begin{align}
\sum_{bc} \epsilon_{abc} ( \partial_c A^c)_{i} = \frac{2\pi }{p} \rho^{l_a}_i. \label{eq:flux-attach-6}
\end{align}
Invoking the conservation law of the $l_a$-quasiparticles given in Eq. (\ref{eq:ccl1}) we get $\sum_{bc} \epsilon_{abc} (\partial_c \dot{A}^c)_{i}= -(2\pi/p) \sum_{bc} \epsilon_{abc} (\partial_c [j^l]^c)_i$ and the solution is 
\begin{align}
\dot{\bm A}_{i}=-\frac{2\pi}{p} {\bm j}^l_i. \label{eq:flux-attach-6b}
\end{align}
Equations (\ref{eq:flux-attach-5})-(\ref{eq:flux-attach-6b}) formalize the constraints in the X-cube model, which are embodied in the Lagrangian
\begin{align}
\mathcal{L} & =  {\bm E}_i \cdot \partial_t {\bm A}_i +\frac{1}{2}\sum_{abc} |\epsilon_{abc}| A_i^a [j^f]_i^{bc} + \frac{2\pi}{p} {\bm E}_i \cdot  \bm j^l_i \nn
& +\frac{p}{2\pi} E^{a0}_i  \Bigl(\sum_{bc} \epsilon_{abc} \partial_c A^c- \frac{2\pi}{p} {\rho}^{l_a} \Bigl)_{i} \nn
&-A^0_i  \Bigl(\sum_{abc} |\epsilon_{abc}| \frac{1}{2}\partial_a\partial_b E^c-\rho^f \Bigl)_i,
\end{align}
where $E^{a0}_i$ and $A_i^0$ are the Langrange multipliers. After re-defining
\begin{align}
\sum_{a} |\epsilon_{abc}| A^a_i &\rightarrow A^{bc}_i\nn
-(2\pi/p) E^a_i &\rightarrow E^a_i\nn
(\rho^f_{i} ,{\bm j}^f_{i} ) & \rightarrow (-\rho^f_{i} ,-{\bm j}^f_{i}),
\end{align}
and taking continuum limit, we can get Eq. (\ref{eq:XCL}).
%
%
%

\bibliography{SC}
\end{document}